# Managing multi-facet bias in collaborative filtering recommender systems

Samira Vaez Barenji[1], Saeed Farzi[2]
[1]Computer Engineering Faculty, K. N. Toosi University of Technology, Tehran, Iran, samira.vaezbarenji@email.kntu.ac.ir
[2]Computer Engineering Faculty, K. N. Toosi University of Technology, Tehran, Iran, Iran School of Computer Science, Institute for Research in Fundamental Sciences (IPM), Tehran, Iran, saeedfarzi@kntu.ac.ir

## Abstract
Due to the extensive growth of information available online, recommender systems play a more significant role in serving people's interests. Traditional recommender systems mostly use an accuracy-focused approach to produce recommendations. Today's research suggests that this single-dimension approach can lead the system to be biased against a series of items with certain attributes. Biased recommendations across groups of items can endanger the interests of item providers along with causing user dissatisfaction with the system. This study aims to manage a new type of intersectional bias regarding the geographical origin and popularity of items in the output of state-of-the-art collaborative filtering recommender algorithms. We introduce an algorithm called MFAIR, a **m**ulti-**fa**cet post-processing b**i**as mitigation algo**r**ithm to alleviate these biases. Extensive experiments on two real-world datasets of movies and books, enriched with the items' continents of production, show that the proposed algorithm strikes a reasonable balance between accuracy and both types of the mentioned biases. According to the results, our proposed approach outperforms a well-known competitor with no or only a slight loss of efficiency.

## 1. Introduction

Recommender system (RS) is a kind of information filtering system that helps users find their items of interest [1]. These systems predict users' preferences and ratings for unknown items by learning their interactive patterns in historical data. The training data fed to a recommender model may be imbalanced, which would cause the model to discriminate unfairly against individual items or groups of items [2, 3]. Bias occurs when discrimination systematically results in unfair outcomes [4]. Users are among the most important stakeholders in any RS, but they are not the only group [5]. The interests of system owners, item providers, and users are intertwined in these systems. Any kind of bias can endanger the interests of one or more of these groups. For instance, if item providers do not receive enough exposure in recommendations, they would eventually leave the system. On the other hand, since provider bias results in an unbalanced share of visibility for providers, this limits the user's choices and thus results in their dissatisfaction.

Sometimes the origin of bias is directly related to the sensitive attributes of the items (e.g., race or gender) [6, 7]. This might reveal the prevalence of cultural or political discrimination within a society, which is reflected in the data and reinforced by the RS model [8]. These sensitive attributes can belong to item providers or users. For example, users in a movie recommendation system may observe a different distribution of movie genres in their recommendation list depending on their gender [9]. Data imbalance, however, can be tied to the underlying mechanisms of a society or an industry. It does not always point to social discrimination or a biased data collection process. An obvious case of this phenomenon is the modern film industry, where Hollywood holds the highest market share of both movies produced and revenues[1].

Gómez et al. [10] demonstrate how state-of-the-art collaborative filtering (CF) algorithms reinforce the geographic imbalance of providers in the input data. They propose a post-processing method to balance the actual distribution of items belonging to different geographical areas. They suggest two measures of visibility and exposure which we are inspired by to assess the existing geographic bias in recommendations.

Different kinds of bias in information retrieval and RS are considered in the literature, and various definitions are provided for each [11]. In this study, bias in RS is simultaneously examined from two different perspectives: the geographical origin of item providers and the items' popularity. Each of these two cases has been studied independently in recent studies [3, 12, 13]. But when items are discriminated against in several different ways, the problem of intersectional bias arises. Intersectional bias means that an under-represented item is likely to be subject to more discrimination than other under-represented items due to having more than one sensitive attribute at a time [14]. Therefore, bias management for a single aspect cannot ensure the fairness of the system.

In this research, the goal is to simultaneously improve two facets of bias in CF recommender systems to overcome the intersectional bias problem. In order to examine this problem in the state-of-the-art CF algorithms, we consider two well-known domains of RS: movies and books. A continent is considered as the granularity level of geographical origin. In both of our datasets, North America accounts for the largest share in terms of both produced items and given ratings. As shown in [10], this imbalance leads to biased recommendations across continents. According to UNESCO, our two study areas, cinema and literature are effective platforms for culture, education, entertainment, and propaganda [3]. Therefore, assessing how recommender systems might encourage a nation's product consumption is an important topic in terms of provider fairness [3].

Furthermore, the items from both datasets are classified into different popularity groups. Similar to geographic imbalance, popularity imbalance also leads to unfair results for long-tail or under-represented groups [15, 16]. The overall objective is to provide a better opportunity for under-represented items to be exposed by improving fairness from two different aspects, which in turn can attract the user's interest.

Our proposed algorithm conducts a set of optimal swaps in recommendation lists to improve the distribution of under-represented groups in the output of a recommender algorithm. These swaps slightly demote the over-represented items in the lists while promoting the under-represented items with the highest scores. Note that the geographical distribution of recommendations is adjusted according to the continents' distribution in the input data, based on the equity concept [17]. While establishing geographic equity, the algorithm gains more visibility and exposure [3] for items from sensitive popularity groups.

We intend to examine both mentioned biases in terms of visibility and exposure and provide an algorithm to manage them. Additionally, comparing our approach to the one proposed in [3] demonstrates how it ignores popularity bias and occasionally even exacerbates it. The results of our experiments show that the presented approach in this paper can establish a reasonable and flexible balance between the two facets of bias while maintaining the quality of the recommendations. To the best of our knowledge, there are very few studies of intersectional bias in the literature, and a great lack is felt in the scope of the multifaceted bias examination.

---

[1] https://www.boxofficepro.com/global-box-office-rebounds-to-21-3-billion-in-2021-as-exhibition-transitions-from-closures-to-blockbusters/

The rest of the paper is structured as follows: in Section 2, we present background and related work. The problem formulation and our proposed approach are discussed in Section 3, and in Section 4, experimental studies are provided. We conclude the paper and suggest future work in Section 5.

## 2. Background and related work

In this section, previous works in the literature are discussed in two separate parts. The first part reviews some recent studies in the traditional RS area. The second part focuses on fairness in recommendation and ranking, which is the main scope of this work.

### 2.1. Traditional recommender systems

The foundation of traditional RSs is solely built on maximizing accuracy and user satisfaction [18, 19]. They use the user's previous interactions and preferences to predict their actual interests. Reddy et al. [20] present a content-based movie recommender system that operates based on genre correlation. It recommends those movies to the user that are likely to be of their favorite genre. Jiang et al. [21] propose a CF recommender system in the E-commerce domain that uses trust data to achieve higher accuracy. A cross-domain CF algorithm is proposed in [22], which expands user and item features via the latent factor space of auxiliary domains to address the data sparsity problem of CF methods. Sujithra Alias Kanmani et al. [23] use the temporal characteristics of transactions and demographic attributes to solve the cold start and data sparsity problems as well as increase accuracy in a movie recommender system. CF methods are more prone to bias generation since they leverage the existing patterns of user interactions in historical data to make recommendations. Furthermore, we don't need descriptive features of items or demographic attributes about users to implement CF methods. Therefore, CF methods will be the main focus of this study.

### 2.2. Fairness in recommendation

A recommender system, like any data-driven system, is prone to amplifying the inherent bias in data and delivering unfair output. The fairness issue in RS has attracted a lot of attention in recent years. The literature has focused on either user fairness, provider fairness, or both.

Burke et al. [24] propose a balanced neighborhood mechanism to maintain personalization in RS while improving the fairness of recommendations for protected groups. Li et al. [25] classify users into different groups based on their activity level in the system, and by offering a re-ranking technique, they reduce the bias of the RS towards a protected group. Abdollahpouri et al. [26] examine popularity bias from a user perspective by dividing users into different groups according to their interest in long-tail items. Unlike them, we measure popularity bias based on the distribution of popularity groups in the entire input data rather than considering a personalized form of popularity bias for users. Managing geographic bias along with a personalized version of popularity bias for users can be addressed in future work.

Regarding provider-side fairness, Gómez et al. [3, 10] adjust the geographic distribution of items in the movie and book recommender systems to establish the geographic equity of providers in the output. They propose a post-processing method to achieve this goal. They also applied their method to an educational RS [12]. We demonstrate how popularity bias is ignored in their work and then combine geographic and popularity fairness to achieve fair and high-quality recommendations. Qi et al. [27] introduce a fairness-aware news recommendation framework that learns fair recommendation models for various news provider groups from biased user data. Abdollahpouri et al. [28] present a configurable regularization-based framework to establish a flexible trade-off between accuracy and coverage rate of long-tail items in a learning-to-rank RS. Zhang et al. [13] leverage the popularity bias in recommendation using a new training paradigm for recommendation, which adjusts the recommendation score with the desired popularity bias via causal intervention. Lin et al. [29] propose measures and a framework to assess and manage popularity bias in conversational recommender systems.

Considering both item-side and user-side fairness, Ranjbar Kermany et al. [16] present a fairness-aware multi-objective recommender system using an evolutionary method. They consider personalized diversity and popularity along with provider fairness to achieve high-quality recommendations. The two-sided fairness problem in E-commerce platforms is also formulated as a multi-objective optimization problem in [30]. Patro et al. [31] map the fairness-aware recommendation problem to the constrained fair allocation of indivisible goods to ensure a minimum level of fairness for both users and providers. In this research, we will only focus on provider fairness and popularity bias while maintaining the accuracy of recommendations.

## 3. Proposed method

In this Section, we first define the preliminaries of the problem as well as both popularity and geographic bias. Then, to address the issue, the post-processing mitigation approach is presented in Section 3.2.

### 3.1. Preliminaries

Consider a set of users $U = \{u_1, u_2, \dots u_n\}$ and a set of items $I = \{i_1, i_2, \dots, i_m\}$ where each user has rated one or more items. The whole data can be formed as a set of triplets in which each record consists of user ID, movie ID, and the corresponding rating. To train and evaluate the recommender model, the data is split into test and train sets in a fixed ratio. The whole training data can be defined as $D = \{(u, i, r_{ui}) : u \in U, i \in \Phi_u, \Phi_u \subseteq I\}$. $r_{ui}$ denotes the rating of user *u* for item *i*, and $\Phi_u$ represents the subset of items rated by user $u$ in the input data. After feeding the algorithm with the training dataset, the recommender learns a function that estimates the relevance value $\widehat{r_{ui}}$ for each user-item pair $(u, i)$ that doesn't exist in the training data ($(u, i, \widehat{r_{ui}}) \notin D$). The top-$n$ items with the highest predicted relevance values $\widehat{r_{ui}}$ form their recommendation list $\Psi_u$. In the following, two concepts of popularity and geographic bias are explained based on the above notations.

**Popularity bias:** The popularity of an item is defined as the number of ratings the item has received in the dataset [13, 32]. Sorting the items based on their popularity rate in the input data, they fall into three separate categories, with 10% of the items on top of the list in group 1 (the most popular), the next 10% in group 2 (relatively popular), and the remaining items being placed in group 3 (unpopular). Since the frequency of the items in these groups is very unbalanced, the total popularity value of a group is not a suitable measure to determine the share of that group's popularity in the whole data. According to equation 1, the total popularity value of each group is divided by the total number of items in that group.

$$P_g = \frac{\sum_{i \in g} \pi_i}{|g|} \tag{1}$$

Where $g$ is a popularity group, and $\pi_i$ is the popularity of item $i$. The $P_g$ values are then normalized to sum the popularity values of the three groups to 1. As shown in Figure 1, there is a severe imbalance across the popularity proportion of popularity groups. Our goal is to bring the share of each popularity group in the whole recommendation lists $\Psi$ close to the value obtained for that group in equation 1, such that the efficiency of the system does not decrease as much as is feasible.

**Geographic bias:** As mentioned in Section 1, the production continent serves as the geographic granularity level in this study. Figure 2 illustrates the distribution of produced items and given ratings at the continent level for both the MovieLens-1M and Book-Crossing datasets. As shown in the figure, there is a significant imbalance in the geographical distribution of items and ratings in both datasets.

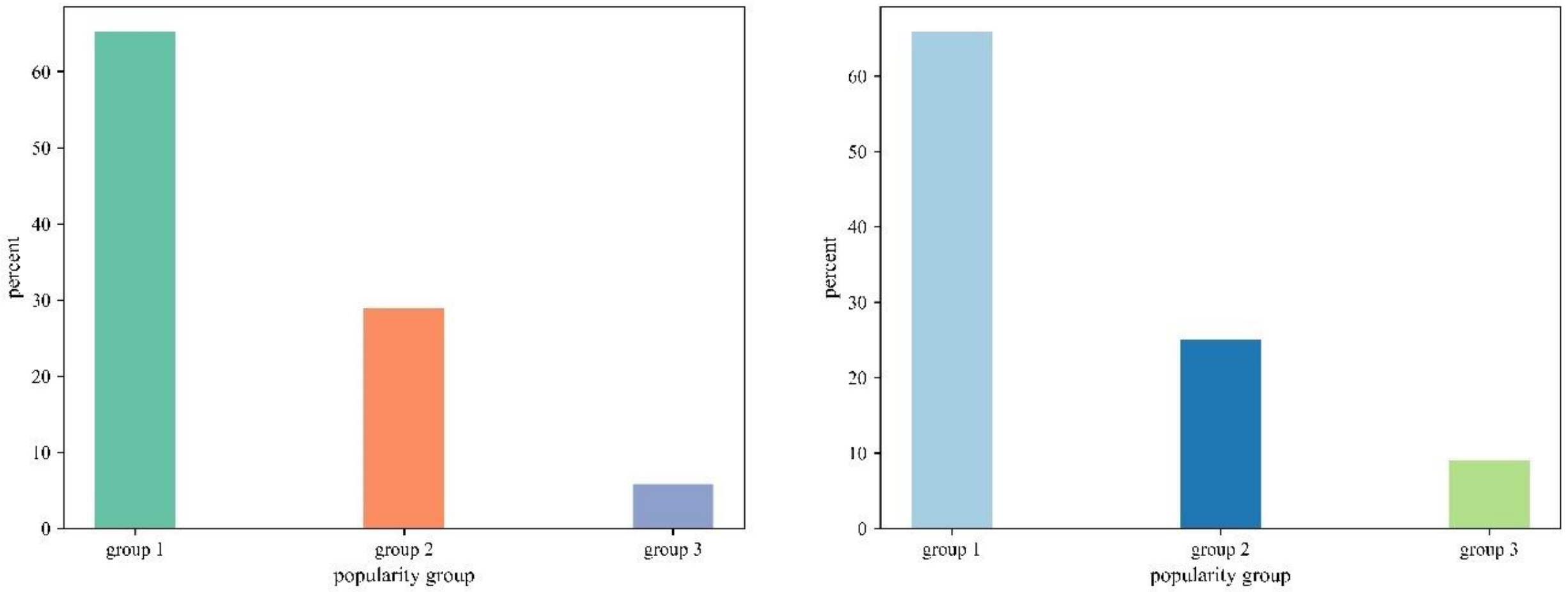

(a) MovieLens-1M (b) Book-Crossing

Figure 1. Popularity representation in the input data: a) MovieLens-1M and b) Book-Crossing datasets. Group 1 includes the most popular items. Groups 2 and 3 include long-tail items.

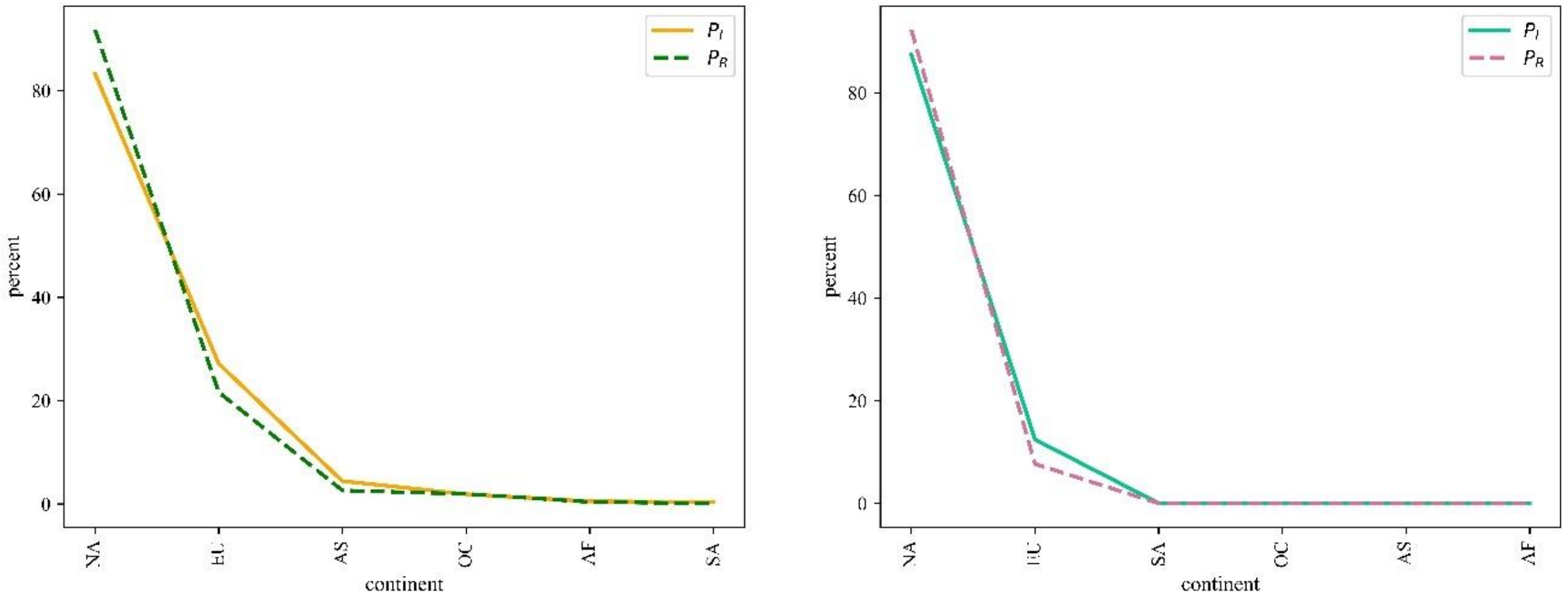

(a) MovieLens-1M (b) Book-Crossing

Figure 2. Continent representation in the input data: a) MovieLens-1M and b) Book-Crossing datasets. $P_I$ denotes the proportion of items belonging to a continent, and $P_R$ denotes the proportion of ratings given to items of a continent.

In order to establish geographical equity in the recommendations, the distribution of the producer continents of items in $\Psi$ is expected to closely match their distribution in the input data, which is henceforth called the target proportions. The target proportions can be defined in two ways: *i*) The proportion of items belonging to each continent in the input data given in equation 2. *ii*) The proportion of available ratings for items of each continent in the input data given in equation 3.

$$P_I(c) = \frac{|\{i: i \in \Phi^c\}|}{\sum_{c \in C} |\{i: i \in \Phi^c\}|} \quad (2)$$

$$P_R(c) = \frac{|\{r_{ui}: u \in U,\ i \in \Phi_u^c\}|}{\sum_{c \in C}|\{r_{ui}: u \in U,\ i \in \Phi_u^c\}|} \quad (3)$$

$\Phi_u^c$ denotes the items belonging to continent $c$, which are rated by user $u$. $P_I(c)$ and $P_R(c)$ are the item-based and rating-based target proportions for continent $c$, respectively. Note that, in our study, an item can belong to more than one continent. The goal is to match the distribution of continents in $\Psi$ with one of the target proportions mentioned above while imposing the least loss in accuracy.

Figure 3 illustrates a toy example where the output of a vanilla (i.e., unaware of bias) traditional CF technique is compared to two different fairness-ware recommendation lists.

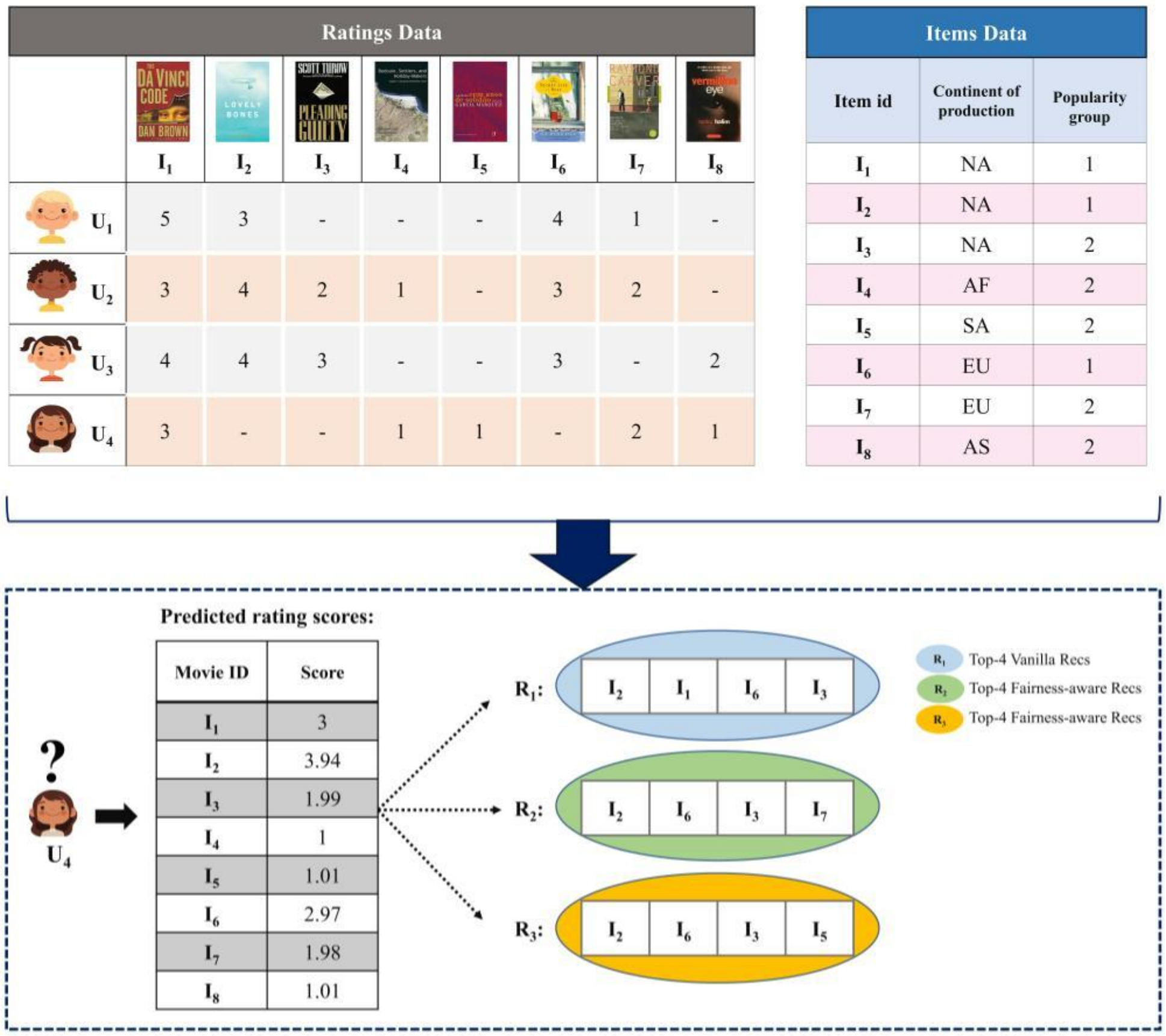

Figure 3. A toy example to explain the intersectional bias of popularity and geographical origin. $R_2$ accounts for geographic bias, and $R_3$ accounts for both geographic and popularity bias.

As shown in the figure, there is a geographical imbalance in the item dataset. Out of the total of eight books, NA (North America) and EU (Europe) represent three and two of them, respectively. While AF (Africa), SA (South America), and AS (Asia) are each represented by a single item. Additionally, in the rating dataset, some items like $I_4$ or $I_5$ have received fewer ratings than others, or the average rating value they have received is quite low. As a result, items $I_3$, $I_4$, $I_5$, $I_7$, and $I_8$ are placed in the second popularity group (i.e., items with relative popularity). Items $I_4$, $I_5$, and $I_8$ are subject to intersectional discrimination since they belong to an under-represented continent, and at the same time, to the second popularity group. There are three recommendation lists for user 4 at the bottom of the figure. $R_1$ includes the top-4 recommendations based on the relevance scores predicted by the matrix factorization algorithm.

There is one item from EU and three from NA in $R_1$, as would be expected. Moreover, only one item from the second popularity group is included in this list. $R_1$ is re-ranked in two ways, including $R_2$ and $R_3$, to generate a fairness-aware recommendation list. In $R_2$, $I_1$ is excluded from the list, and the 5$^{th}$ highest-scoring item, according to the predicted scores, is included in the list. $I_1$ is selected as the best choice to be excluded from the list since it is a geographically over-represented item from the first popularity group, and has a lower score than its peers and is expected to result in a lower loss if we exclude it. Similarly, item $I_1$ in $R_3$ is excluded from the list and replaced with the 6$^{th}$ highest-scoring item for the target user. $I_5$ can be a better choice than $I_7$ to be added to the list since it belongs to an under-represented continent that has got no share in the list, and at the same time, belongs to the second popularity group. As a result, it experiences a kind of intersectional discrimination. On the other hand, there is just a slight difference in their predicted scores with $I_7$, which leads to a negligible efficiency loss. $R_3$ has intelligently handled both kinds of geographic and popularity bias overall, highlighting our intended approach in this study. In essence, both types of biases will be measured based on a target distribution derived from the input data. We refer to the set of recommended items to user $u$ belonging to a group $g$ (i.e., a continent or popularity group) as $\Psi_u^g$. In addition, the measurement of both biases will be based on two scales: visibility and exposure [3, 33]. Visibility means the proportion of items from each group in the top-$k$. Exposure describes how far items in a group have been placed in higher positions. Finally, all the notations are summarized in Table 1.

Table 1. A summary of notations.

| **notation** | **meaning** |
|---|---|
| U | Set of users |
| I | Set of items |
| $D$ | Training data |
| $\Phi_u$ | Subset of items rated by user $u$ in the training data |
| $\widehat{r_{ui}}$ | Predicted relevance value for the user-item pair ($u$, $i$) |
| $\Psi_u$ | Recommendation list of user $u$ |
| $\Psi_u^g$ | Set of items belonging to group $g$ (e.g., a continent or popularity group) in $\Psi_u$ |
| $\Psi$ | Total recommendation lists |
| $P_g$ | Target proportion for popularity group $g$ |
| $P_I(c)$ | Item-based target proportion for continent $c$ |
| $P_R(c)$ | Rating-based target proportion for continent $c$ |

### 3.2. Approach

In this section, we introduce MFAIR, a **m**ulti-**fa**cet post-processing b**i**as mitigation algo**r**ithm. The proposed algorithm adopts a flexible and adjustable approach to managing the trade-off between popularity bias and accuracy. Algorithm 1 provides the pseudo-code for MFAIR. The algorithm operates based on raising the position of an item in the list, considering the geographic and popularity bias in the entire recommendations. An item is chosen to be promoted, considering all the recommendation lists and all possible swaps to impose the least loss on the system. Algorithm 2 presents the supporting function called in MFAIR. Figure 4 shows the main procedure of our approach, considering both visibility and exposure of popularity groups and continents. The distribution of groups is adjusted in two phases. The first phase re-ranks the lists to mitigate bias in terms of visibility, and the second phase in terms of exposure. In both phases MFAIR takes the recommendation lists for all users, the target proportions for popularity groups and continents, as well as the bias type (i.e., visibility or exposure) as input. In the second phase, the input recommendation lists are the output of the previous phase, re-ranked for visibility. This two-phase multi-facet bias mitigation process results in fairness-aware recommendation lists in terms of both visibility and exposure.

MFAIR adjusts the visibility or exposure of popularity groups and continents in recommendations, taking into account their bias rates. The bias rate of a group is the difference between the target proportion of the group and its actual exposure or visibility proportion in the recommendations. First, the required data structures are initialized, and the existing biases in the recommendations are calculated for popularity groups and continents (lines 2 and 3) based on the bias type.

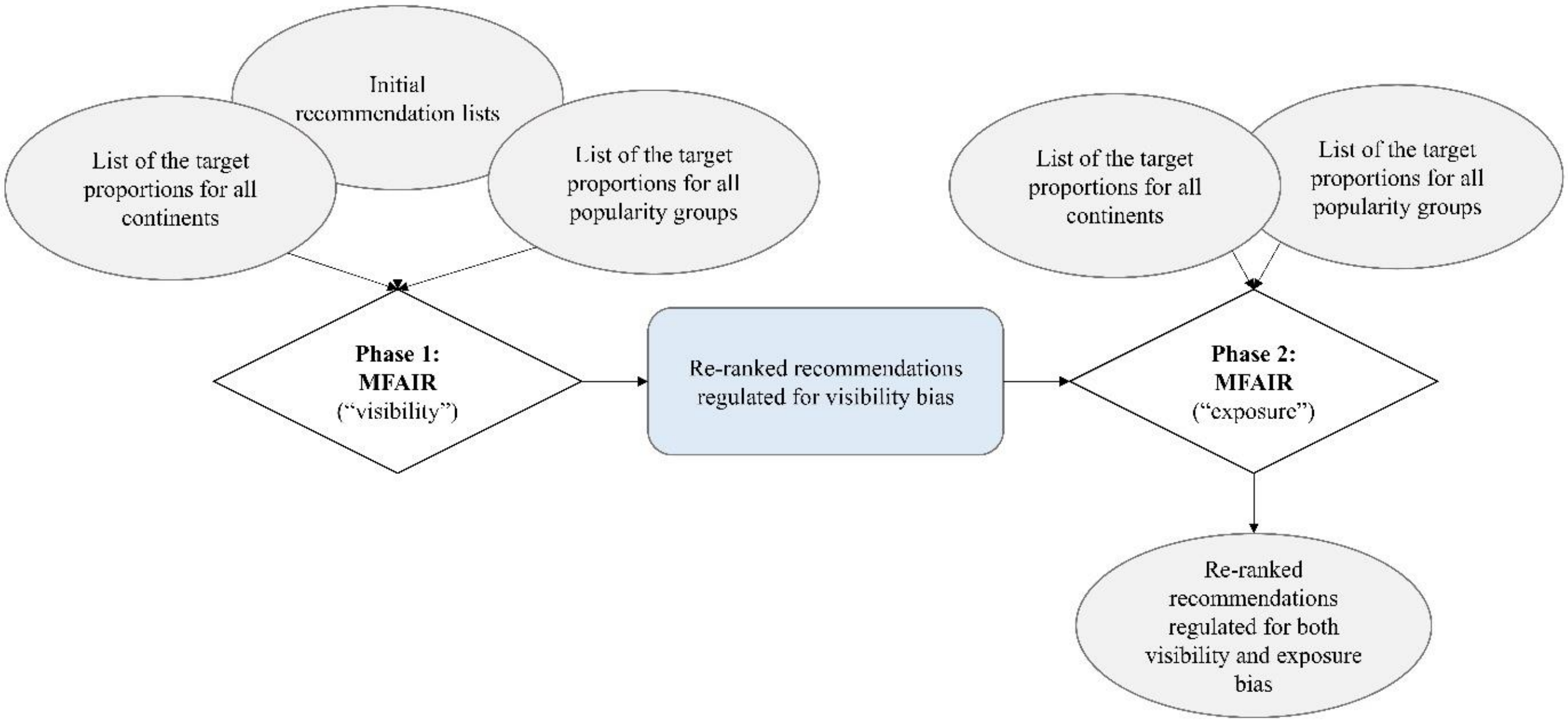

Figure 4. A flowchart of the proposed approach.

**ALGORITHM 1. MFAIR: MULTI-FACET BIAS MITIGATION ALGORITHM**

```
Input:      recs: initial recommendation lists
            targetCont: list of the target proportions for all continents
            targetPop: list of the target proportions for all popularity groups
            biasType: visibility or exposure
Output:     recs: re-ranked fair recommendation lists
define MFAIR (recs, targetCont, targetPop, biasType)
begin
    recsUp, recsDown, swaps ← list (), list (), list ();
    contDelta, popDelta ← calculate biases for continents and popularity groups based on biasType;
    foreach user ∈ recs do
        foreach rec ∈ user.recs (top-n) do //  loop through the top-n recommended items to each user
            if rec.position == True and underRepresented (rec, biasType ) == False then
                recsDown.add (rec);
            elseif rec.position == False and underRepresented (rec, biasType ) == True then
                recsUp.add (rec);
            end
        end
        while recsUp.size != 0 and recsDown.size != 0 do
            recUp ← recsUp.pop(first);
            recDown ← recsDown.pop(last);
            loss = recDown.relevance – recUp.relevance;
            swaps.add (id, recDown, recUp, loss);
        end
    end
    addPenalty (swaps, popDelta); // Add some penalty to each loss value based on the popularity group of items
    swaps.sortAscending (key = loss); // sort possible swaps based on the loss values
    while any (contDelta > 0) and swaps.size != 0  do
        swap ← swaps.pop (nextItem);
        if swap.position == True and underRepresented (swap, biasType) == True then
            recs.swap ( swap.recDown,  swap.recUp);
            contDelta, popDelta ← update biases for continents and popularity groups based on biasType;
        end
    end
    return recs;
end
```

**ALGORITHM 2. SUPPORTING FUNCTION**

define **addPenalty** (*swaps, popDelta*)
**begin**
    averageLoss = $\overline{|swaps.loss|}$; // calculate the average of absolute loss values
    **foreach** *swap* ∈ *swaps* **do**
        **if** *popDelta* [*swap*.recUp] > 0 and *popDelta* [*swap*.recDown] < 0 **then**
            *swap*.loss -= (*averageLoss* * *eps*);
        **elseif** *popDelta* [*swap*.recUp] < 0 and *popDelta* [*swap*.recDown] > 0 **then**
            *swap*.loss += (*averageLoss* * *eps*);
        **end**
    **end**
    **return** *swaps*
**end**

Then two lists of items that can be demoted and promoted are created as *recsDown* and *recsUp*, respectively. *recsDown* contains items that don't belong to an under-represented continent and can be demoted to lower positions in the list. *recsUp* includes items that belong to an under-represented continent, and the geographic bias can be reduced by promoting their position in the list. *Underrepresented(.)* returns true if the given item is under-represented in *recs*.

In lines 12-17, the pair of items are stored in *swaps* whose swapping in a user's list causes the minimum loss. The selection process of these pairs is as follows: At each iteration, the item with the lowest score from *recsDown* and the item with the highest score from *recsUp* is selected. The two selected items represent a possible minimum-loss swap, and they are added as a record to *swaps* along with the potential loss value that would result from making this swap. In line 19, a penalty value is added to each swap loss to prioritize swaps based on their imposable popularity bias. Then, *swaps* is sorted based on the loss values, and they are performed respectively until we reach the desired geographical distribution. Finally, the re-ranked lists are returned.

Algorithm 2 first calculates the average of absolute loss values as the base value of the penalty. Then, for each swap, it adds or subtracts an epsilon of the calculated average to or from each loss value. If the item to be swapped out is in an over-represented popularity group and the one to be swapped in is from an under-represented popularity group, the penalty value will be subtracted from its loss to prioritize this swap, and in the opposite case, the value will be added to its loss. In case the swapping items are both from an under- or over-represented group, the loss remains unchanged. *eps* is an adjustable parameter with a value between 0 and 1. If it is equal to 1, the popularity bias will improve the most.

## 4. Experimental studies

We perform a set of experiments on two real-world datasets enriched with producer continents of items to evaluate our proposed approach. The Java LibRec library v.2 is used to implement some state-of-the-art CF recommender algorithms and generate initial recommendation lists. The length of each recommendation list (top-*n*) will be 150, while the re-reranked lists will contain 20 items (top-*k*) for each user. In the following, both datasets of our study are introduced. Then, the evaluation measures are described in section 4.2. Finally, we report the results of our experiments and compare our approach to a state-of-the-art post-processing method in this scope.

### 4.1. Datasets

**MovieLens-1M.** This is a well-known movie dataset in the recommendation domain provided by the GroupLens research group[2]. It contains 1,000,209 ratings given by 6,040 users on 3,706 movies. Each user has rated at least 20 movies in the range (1-5). The IMDb ID provided for each movie allowed us to enrich the dataset with movies' continents of production using OMDB APIs[3]. The items in the dataset belong to one or more of the six continents of Africa, Asia, Europe, North America, Oceania, and South America.

**Book-Crossing.** The dataset is collected by Cai-Nicolas Ziegler over a 4-week crawl from the Book-Crossing community [34]. It includes 1,149,781 ratings given by 105,283 users on 340,556 books in the range (1-10). Since the rating data was quite sparse for this dataset, we removed users and books with less than five ratings. This process resulted in 132,762 ratings given by 7,704 users on 13,631 items. We enriched the dataset with the books' continents of production using the data provided by [3]. The books in the dataset belong to one of the four continents of Europe, North America, Oceania, and South America.
Table 2 provides an overview of statistics for both datasets.

Table 2. A summary of statistical information of both datasets: Number of items ($|\boldsymbol{I}|$), number of users ($|\boldsymbol{U}|$), number of ratings ($|\boldsymbol{R}|$), and item's proportion ($\boldsymbol{P_I(c)}$) – ratings' proportion ($\boldsymbol{P_R(c)}$) – average rating ($\boldsymbol{AR(c)}$) across continent *c* (AF: Africa, AS: Asia, EU: Europe, NA: North America, OC: Oceania, SA: South America)

| Dataset | $\vert\boldsymbol{I}\vert$ | $\vert\boldsymbol{U}\vert$ | $\vert\boldsymbol{R}\vert$ | $\boldsymbol{P_I(c)}$- $\boldsymbol{P_R(c)}$- $\boldsymbol{AR(c)}$ |
|---|---|---|---|---|
| **MovieLens-1M** | 3,706 | 6,040 | 1,000,209 | AF : 0.42- 0.32- 3.97<br>AS : 3.72- 2.24- 3.45<br>EU : 23.25- 18.22- 3.66<br>NA : 70.66- 77.55- 3.56<br>OC : 1.62- 1.60- 3.65<br>SA : 0.35- 0.08- 3.21 |
| **Book-Crossing** | 13,631 | 7,704 | 132,762 | EU : 12.68- 7.70- 7.83<br>NA : 87.21- 92.27- 7.80<br>OC : 0.07- 0.02- 7.78<br>SA : 0.04- 0.02- 8.41 |

$\boldsymbol{P_I(c)}$ and $\boldsymbol{P_R(c)}$ are in percentage.

### 4.2. Evaluation metrics

As mentioned in the previous sections, the bias evaluation in this study will be in two dimensions. In the following, the evaluation measures for both dimensions of visibility and exposure are explained.
**Visibility Bias (VB).** The visibility bias for a continent *c* is the difference between the target proportion (rating-based or item-based) of *c* and its actual visibility proportion in the top-*k* items of recommendation lists, formulated as follows:

[2] https://grouplens.org/
[3] http://www.omdbapi.com/

$$VB(c)@k = \left( \sum_{u \in U} \frac{|\Psi_u^c @k|}{\sum_{c \in C} |\Psi_u^c @k|} \right) - P_{I/R}(c) \quad (4)$$

Where $C$ is the set of all available continents in the recommendation list of user $u$, and $\Psi_u^c @k$ denotes the top-$k$ recommendations of user $u$ that belong to continent $c$.

The actual proportion of a popularity group in the top-$k$ recommendations is calculated slightly differently than for geographical groups. Thus, the visibility bias for a popularity group $g$ is formulated according to equation 5.

$$VB(g)@k = \left( \frac{V_g @k}{\sum_{g \in G} V_g @k} \right) - P_g \qquad V_g @k = \sum_{u \in U} \frac{|\Psi_u^g @k|}{|\Phi^g|} \quad (5)$$

Where $V_g @k$ is the visibility of group $g$ in the top-$k$ recommendations.

**Exposure Bias (EB).** The exposure bias for a continent $c$ is the difference between the target proportion (rating-based or item-based) of $c$ and its actual exposure share in top-$k$ of the whole recommendation lists, formulated as follows:

$$EB(c)@k = \left( \sum_{u \in U} \frac{\sum_{i \in \Psi_u^c @k} \frac{1}{\log_2(1 + pos_{ui})}}{\sum_{c \in C} \sum_{i \in \Psi_u^c @k} \frac{1}{\log_2(1 + pos_{ui})}} \right) - P_{I/R}(c) \quad (6)$$

$pos_{ui}$ indicates the position of item $i$ in $\Psi_u$. Furthermore, the exposure bias for a popularity group $g$ is formulated according to equation 7.

$$EB(g)@k = \left( \frac{E_g @k}{\sum_{g \in G} E_g @k} \right) - P_g \qquad E_g @k = \sum_{u \in U} \frac{\sum_{i \in \Psi_u^g @k} \frac{1}{\log_2(1 + pos_{\widehat{r_{ui}}})}}{|\Phi^g|} \quad (7)$$

Where $E_g @k$ is the exposure of group $g$ in top-$k$ recommendations.

The positive and negative values of the bias measures indicate, respectively, the over- and under-representation of a group in recommendations. In the best scenario, its value will be zero, suggesting that the distribution of a group in the output is fair (i.e., equal to the target proportion of the group).

Moreover, the *Normalized Discounted Cumulative Gain* (*NDCG*) [35] is used to assess the accuracy of recommendations.

## 4.4. Configuration and setting parameters

All the algorithms are implemented in Python and run on an Intel Core i7-8550U 2.20 GHz CPU with 16 GB RAM. The epsilon parameter in MFAIR is set to 1 for all experiments. This parameter can be tuned to balance the popularity bias rate in recommendations. Decreasing its value can slightly increase the popularity bias, which can, in turn, lead to an improvement in other measures, including accuracy[4].

## 4.5. Mitigation Results

In this section, the results of applying MFAIR to the movies and books datasets are examined in terms of accuracy and bias. Figures 5 and 6 show the results of our experiments on the movies and books datasets, respectively. The bar graphs and scatter plots represent the geographic and popularity biases, respectively. As shown by the thick bars in the figures provided for the two datasets, the geographic bias rate of original recommender algorithms in rating-based experiments is often lower than that of an item-based experiment. In other words, the algorithms can better adapt to the geographical distribution of user interactions rather than the distribution of items.

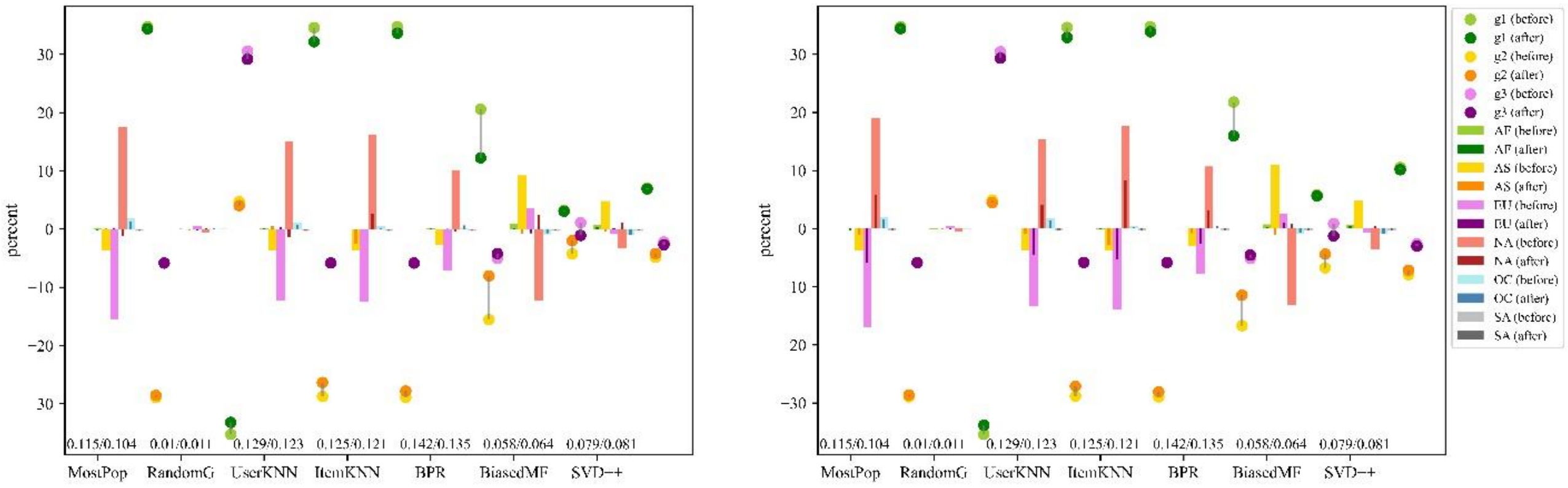

(a) Visibility Disparity (Item-based) (b) Exposure Disparity (Item-based)

(c) Visibility Disparity (Rating-based) (d) Exposure Disparity (Rating-based)

Figure 5. **Bias representation in MovieLens-1M**. Visibility and exposure bias of the state-of-the-art CF algorithms on movies dataset based on $P_I$ (figures a and b) and $P_R$ (figures c and d). The bar graphs and scatter plots represent the biases for continents and popularity groups, respectively. Thick bars show the bias in vanilla recommendations, and thin bars indicate the bias after applying MFAIR in two phases. The text at the bottom of each figure presents the resulting NDCG of each algorithm before and after mitigation, separated by '/'.

[4] We performed extensive experiments with different values of epsilon. To contextualize our approach and compare it to the baseline one, we considered this parameter equal to 1 in all experiments.

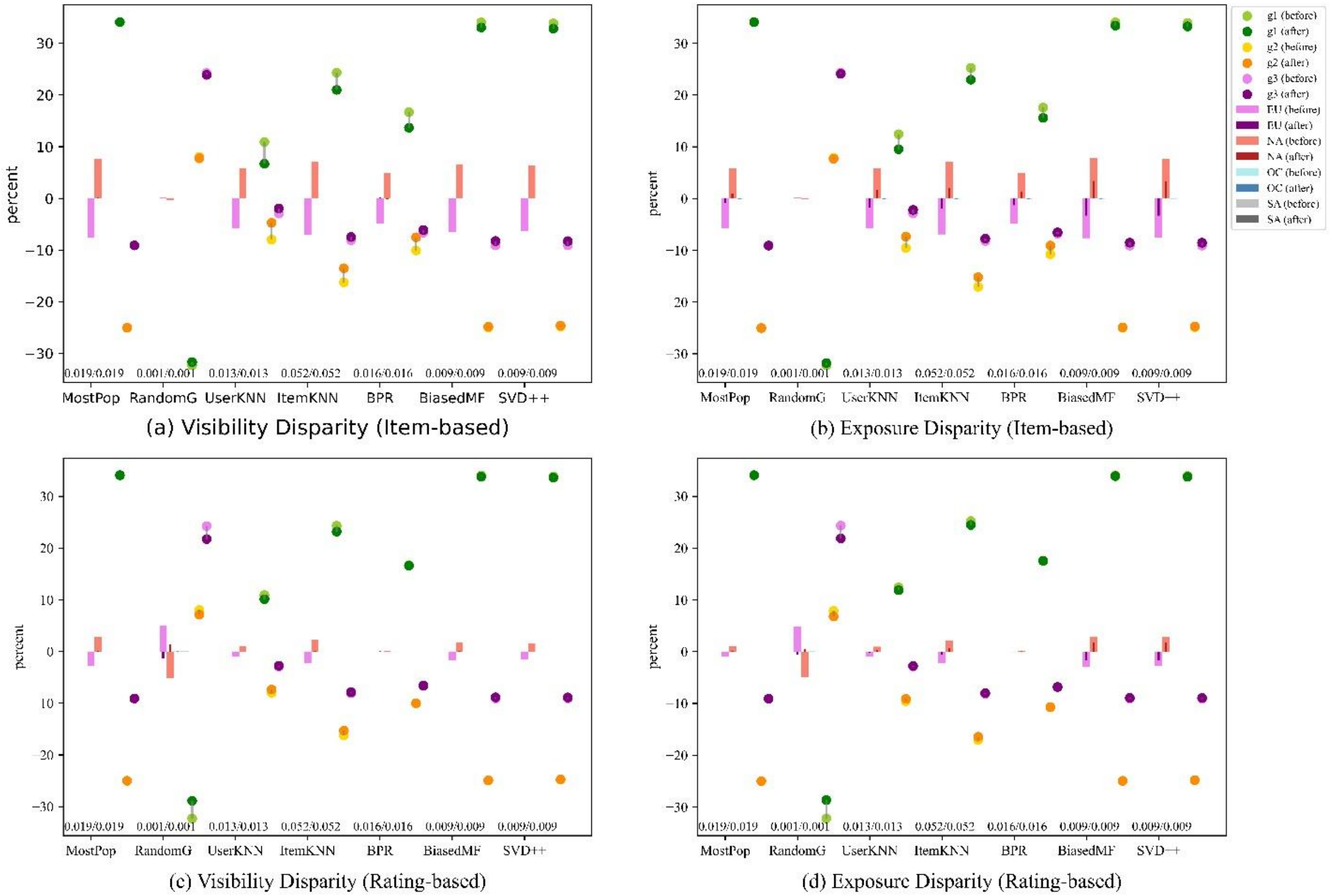

Figure 6. **Bias representation in Book-Crossing**. Visibility and exposure bias of the state-of-the-art CF algorithms on books dataset based on $P_I$ (figures a and b) and $P_R$ (figures c and d). The bar graphs and scatter plots represent the bias for continents and popularity groups, respectively. Thick bars show the bias in vanilla recommendations, and thin bars indicate the bias after applying MFAIR in two phases. The text at the bottom of each figure presents the resulting NDCG of each algorithm before and after mitigation, separated by '/'.

This case is only violated in RandomGuess, which fits better with the geographical distribution of items by recommending random items. However, in terms of NDCG, this algorithm is the least efficient technique. As shown in the figures, the height of the thin bars in both datasets, which show the results produced by MFAIR, is much lower than the corresponding thick bars. Even in some cases, especially in the books dataset, it is almost zero. Popularity bias has also improved after mitigation across most experiments. The results show that in both datasets, the popularity bias has improved to a higher degree in the item-based mitigation than in the rating-based scenario since, in the former evaluation, the popularity of items in user interactions is not taken into account. All the recommender algorithms except RandomGuess over-represent the first popularity group while under-representing the other groups.

Looking more closely at the performance of algorithms on each dataset, Figure 5 shows that BPR has resulted in the lowest amount of rating-based geographic bias on the movies dataset, both in terms of visibility and exposure. Additionally, this algorithm has achieved the highest NDCG value of all the others, even after bias mitigation. Therefore, it can be claimed that BPR performs the best in terms of both geographic bias and NDCG, considering the rating-based distribution of movies. On the other hand,

BiasedMF produces the best results in terms of popularity bias in all experiments on the movies dataset. However, in some experiments, the resulting popularity bias of some algorithms is slightly worsened due to geographical regulation. This phenomenon is observed in rating-based experiments with BiasedMF and SVD++. Compared to others, these two algorithms have caused a higher degree of over-representation for Asia and Europe while simultaneously under-representing North America to a higher degree, and since there are more popular items in the North American item group than in other continent groups, mitigating geographic bias will inevitably result in a slightly more popularity bias. But, as we will see in the following, our approach is better optimized than the competitor one in terms of popularity bias. The NDCG values of RandomGuess, BiasedMF, and SVD++ have slightly improved after bias mitigation in both rating- and item-based experiments. The other algorithms experience just a minor or no loss of accuracy after mitigation.

On the books dataset, considering the rating-based experiments, BPR is still the best-performing algorithm in terms of geographic bias. It also has the second-best performance in terms of popularity bias among all other algorithms. The lowest amount of popularity bias is produced by UserKNN. However, itemKNN outperforms the other algorithms in terms of NDCG. Moreover, since the geographical distribution of items in this dataset is highly unbalanced (i.e., a high proportion of items belong to one of the continents of Europe or North America), rating-based experiments have resulted in a lower degree of geographic bias and less geographic bias overall is produced when compared to the movies dataset. Finally, in all of the experiments on this dataset, the NDCG values remain the same after applying the mitigation algorithm.

A remarkable trend in both datasets is that among the two most frequent continents in the input data (i.e., North America and Europe), when one is over- or under-represented, the other is directly affected in the opposite direction (e.g., when one is over-represented, the other gets under-represented). This trend is evident in almost all experiments and in all algorithms. Furthermore, the lowest bias rate in each experiment belongs to highly under-represented continents such as Africa and Oceania. Since the share of these continents in the input data is almost equal to zero which results in the least bias.

In the following, we compare the results of the approach presented in this paper with those of a state-of-the-art competitor method presented in [3]. Their method mitigates the geographic bias in the output of a vanilla recommender algorithm through a re-ranking approach but ignores the popularity bias in the recommendations.

Tables 3-6 show the results of the comparison of the two approaches, MFAIR and the **basel**ine (baseL.), on the movies and books datasets[5]. For both baseline method and MFAIR, both visibility and exposure biases are reported considering the item-based target distribution ($VB_I$ and $EB_I$ rows) and rating-based target distribution ($VB_R$ and $EB_R$ rows) across all continents (Tables 3 and 4) or popularity groups (Tables 5 and 6). Total BS represents the total absolute **b**ia**s** across all groups in a row. For each algorithm, two NDCG values is reported, which are obtained from item- and rating-based re-raking, respectively (from top to bottom). The numbers in bold show an improvement or no change in a measure compared to the one produced by the baseline method.

Tables 3 and 4 show the resulting geographic biases of our experiments. Based on the algorithm and other experimental settings, as the popularity bias decreases, the geographic bias of MFAIR gets slightly better or worse than the baseline's. For example, according to the total BS of algorithms SVD++, RandomGuess, and BiasedMF (in rating-based case), our method has performed better in improving the geographic bias, but in cases of UserKNN and ItemKNN (both in rating-based cases), the competitor performs slightly better while the cost of this improvement is a higher rate of popularity bias. The direct or inverse relationship between geographic and popularity bias is entirely dependent on the algorithm, dataset, target proportions (rating- or item-based), and type of bias measure (exposure or visibility). Additionally, there is no significant difference in terms of NDCG between the two algorithms, and in most experiments, this measure is almost the same or even better with our approach.

[5] All numbers in Tables 3-6 except NDCGs are in percentage.

Tables 5 and 6 show that in both rating- and item-based experiments for all algorithms, the resulting total popularity bias of MFAIR is lower than that of the baseline one. The popularity bias remains the same only for MostPopular on the books dataset when compared to the competitor due to the lack of long-tail items in the algorithm's top-*n* recommendations.

Table 3. The resulting geographic bias and accuracy of different mitigation approaches on MoviesLens-1M dataset[6].

| Algorithm | | AF | | AS | | EU | | NA | | OC | | SA | | Total BS | | NDCG | |
|---|---|---|---|---|---|---|---|---|---|---|---|---|---|---|---|---|---|
| | | baseL. | MFAIR | baseL. | MFAIR | baseL. | MFAIR | baseL. | MFAIR | baseL. | MFAIR | baseL. | MFAIR | baseL. | MFAIR | baseL. | MFAIR |
| MostPop | $VB_I$ | -0.22 | **-0.22** | 0.21 | 0.22 | 0.23 | **0.23** | -1.22 | **-1.22** | 1.32 | 1.33 | -0.33 | **-0.33** | 3.53 | 3.54 | 0.104 | **0.104** |
| | $EB_I$ | -0.24 | **-0.24** | -0.99 | **-0.98** | -5.92 | **-5.91** | 5.93 | **5.91** | 1.55 | **1.55** | -0.33 | **-0.33** | 14.95 | **14.93** | | |
| | $VB_R$ | -0.1 | -0.11 | 1.74 | 1.75 | 1.23 | **1.23** | -4.06 | -4.08 | 1.27 | 1.29 | -0.08 | **-0.08** | 8.48 | 8.53 | 0.106 | **0.106** |
| | $EB_R$ | -0.13 | **-0.13** | 0.55 | 0.56 | -3.75 | **-3.73** | 1.91 | **1.86** | 1.5 | 1.52 | -0.08 | **-0.08** | 7.91 | **7.88** | | |
| RandomG | $VB_I$ | 0.05 | **0.05** | -0.13 | -0.19 | -0.31 | **-0.19** | 0.33 | **0.21** | 0.05 | 0.1 | 0.02 | **0.01** | 0.89 | **0.76** | 0.011 | **0.011** |
| | $EB_I$ | 0.03 | **0.03** | -0.09 | -0.14 | -0.15 | **-0.08** | 0.16 | **0.09** | 0.04 | 0.09 | 0.02 | **0.01** | 0.49 | **0.45** | | |
| | $VB_R$ | -0.01 | **-0.0** | 0.23 | **0.21** | -1.71 | **-1.59** | 1.38 | **1.27** | -0.01 | **-0.01** | 0.12 | **0.12** | 3.46 | **3.21** | 0.011 | **0.012** |
| | $EB_R$ | 0.01 | **0.01** | 0.35 | **0.33** | -0.71 | **-0.65** | 0.21 | **0.17** | -0.0 | **0.0** | 0.14 | **0.13** | 1.42 | **1.3** | | |
| UserKNN | $VB_I$ | 0.06 | **0.06** | 0.37 | 0.45 | 0.41 | **0.36** | -1.27 | -1.35 | 0.76 | 0.8 | -0.33 | **-0.33** | 3.2 | 3.36 | 0.123 | **0.123** |
| | $EB_I$ | 0.04 | **0.04** | -0.87 | **-0.81** | -4.55 | -4.56 | 4.25 | **4.16** | 1.46 | 1.49 | -0.33 | **-0.33** | 11.5 | **11.39** | | |
| | $VB_R$ | 0.2 | **0.2** | 1.69 | 1.92 | 1.09 | **0.95** | -3.69 | -3.83 | 0.8 | 0.84 | -0.08 | **-0.08** | 7.54 | 7.81 | 0.125 | **0.125** |
| | $EB_R$ | 0.17 | **0.16** | 0.51 | 0.69 | -2.59 | -2.66 | 0.52 | **0.38** | 1.47 | 1.51 | -0.08 | **-0.08** | 5.35 | 5.47 | | |
| ItemKNN | $VB_I$ | -0.08 | **0.0** | -3.14 | **-2.57** | 0.0 | **0.0** | 3.32 | **2.71** | 0.22 | **0.19** | -0.33 | **-0.33** | 7.09 | **5.8** | 0.121 | **0.121** |
| | $EB_I$ | -0.16 | **-0.1** | -3.29 | **-2.88** | -5.35 | **-5.28** | 8.89 | **8.38** | 0.23 | **0.2** | -0.33 | **-0.33** | 18.23 | **17.17** | | |
| | $VB_R$ | 0.4 | **0.4** | 1.06 | 1.12 | 1.1 | **1.06** | -2.6 | -2.61 | 0.11 | 0.12 | -0.08 | **-0.08** | 5.36 | 5.38 | 0.122 | **0.122** |
| | $EB_R$ | 0.25 | **0.25** | 0.23 | 0.28 | -2.84 | -2.86 | 2.3 | **2.28** | 0.14 | **0.14** | -0.08 | **-0.08** | 5.84 | 5.88 | | |
| BPR | $VB_I$ | 0.07 | **0.07** | 0.0 | **0.0** | 0.03 | 0.04 | -0.12 | -0.49 | 0.26 | 0.62 | -0.24 | **-0.24** | 0.71 | 1.46 | 0.137 | 0.135 |
| | $EB_I$ | 0.07 | 0.08 | -0.85 | **-0.78** | -2.92 | **-2.61** | 3.37 | **3.09** | 0.21 | 0.46 | -0.24 | **-0.24** | 8.02 | **7.26** | | |
| | $VB_R$ | 0.14 | **0.13** | 2.29 | 2.54 | -1.76 | -2.01 | -0.64 | **-0.07** | 0.03 | -0.52 | -0.06 | -0.08 | 4.92 | 5.35 | 0.141 | 0.14 |
| | $EB_R$ | 0.15 | **0.15** | 1.26 | 1.53 | -3.02 | -3.14 | 1.62 | 2.05 | 0.06 | -0.5 | -0.07 | -0.08 | 6.17 | 7.44 | | |
| BiasedMF | $VB_I$ | 0.16 | **0.15** | -0.69 | -0.81 | -0.89 | **-0.72** | 2.46 | **2.42** | -0.73 | -0.72 | -0.31 | **-0.31** | 5.24 | **5.14** | 0.065 | 0.064 |
| | $EB_I$ | 0.19 | **0.18** | -1.29 | **-1.13** | 1.23 | **1.13** | 0.82 | **0.78** | -0.64 | -0.65 | -0.31 | **-0.31** | 4.49 | **4.17** | | |
| | $VB_R$ | 0.28 | **0.25** | 1.03 | 1.04 | -1.3 | **-1.25** | 0.72 | 0.8 | -0.69 | -0.77 | -0.03 | **-0.07** | 4.03 | 4.17 | 0.065 | **0.066** |
| | $EB_R$ | 0.34 | **0.31** | 3.33 | **1.5** | 0.55 | **0.14** | -3.64 | **-1.21** | -0.58 | -0.68 | -0.01 | **-0.07** | 8.44 | **3.91** | | |
| SVD++ | $VB_I$ | 0.34 | **0.33** | -0.51 | **-0.36** | 0.14 | **0.09** | 1.28 | **1.19** | -0.94 | **-0.94** | -0.3 | **-0.3** | 3.52 | **3.21** | 0.082 | 0.081 |
| | $EB_I$ | 0.44 | **0.43** | 0.07 | 0.17 | 0.09 | **0.05** | 0.59 | **0.53** | -0.88 | **-0.87** | -0.3 | **-0.3** | 2.36 | **2.35** | | |
| | $VB_R$ | 0.5 | **0.48** | 1.1 | **-0.19** | -1.39 | **-1.06** | 0.68 | 1.78 | -0.82 | -0.93 | -0.07 | **-0.07** | 4.58 | **4.52** | 0.083 | **0.083** |
| | $EB_R$ | 0.59 | **0.57** | 1.89 | **-0.48** | 0.5 | **0.69** | -2.17 | **0.16** | -0.73 | -0.87 | -0.07 | **-0.07** | 5.96 | **2.83** | | |

[6] All numbers in tables 3-6 except NDCGs are in percentage.

Table 4. The resulting geographic bias and accuracy of different mitigation approaches on Book-Crossing dataset.

| Algorithm | | EU | | NA | | OC | | SA | | Total BS | | NDCG | |
|---|---|---|---|---|---|---|---|---|---|---|---|---|---|
| | | baseL. | MFAIR | baseL. | MFAIR | baseL. | MFAIR | baseL. | MFAIR | baseL. | MFAIR | baseL. | MFAIR |
| MostPop | $VB_I$ | 0.0 | **0.0** | 0.11 | **0.11** | -0.07 | **-0.07** | -0.04 | **-0.04** | 0.22 | **0.22** | 0.019 | **0.019** |
| | $EB_I$ | -0.83 | **-0.83** | 0.94 | **0.94** | -0.07 | **-0.07** | -0.04 | **-0.04** | 1.89 | **1.89** | | |
| | $VB_R$ | -0.14 | **-0.14** | 0.18 | **0.18** | -0.02 | **-0.02** | -0.02 | **-0.02** | 0.35 | **0.35** | 0.019 | **0.019** |
| | $EB_R$ | -0.05 | **-0.05** | 0.08 | **0.08** | -0.02 | **-0.02** | -0.02 | **-0.02** | 0.17 | **0.17** | | |
| RandomG | $VB_I$ | -0.0 | **-0.0** | 0.0 | **0.0** | 0.0 | **0.0** | 0.0 | **0.0** | 0.01 | **0.01** | 0.001 | **0.001** |
| | $EB_I$ | -0.02 | -0.03 | 0.01 | 0.02 | 0.01 | **0.01** | 0.0 | **0.0** | 0.04 | 0.05 | | |
| | $VB_R$ | -1.48 | **-1.39** | 1.46 | **1.37** | 0.02 | **0.02** | -0.0 | **-0.0** | 2.96 | **2.78** | 0.001 | **0.001** |
| | $EB_R$ | -0.65 | **-0.59** | 0.62 | **0.56** | 0.03 | **0.03** | 0.0 | **0.0** | 1.3 | **1.19** | | |
| UserKNN | $VB_I$ | 0.0 | **0.0** | 0.04 | **0.04** | -0.04 | **-0.04** | 0.0 | **0.0** | 0.08 | **0.08** | 0.013 | **0.013** |
| | $EB_I$ | -1.7 | **-1.69** | 1.75 | **1.75** | -0.05 | **-0.05** | -0.0 | **-0.0** | 3.5 | **3.5** | | |
| | $VB_R$ | 0.0 | **0.0** | -0.0 | **-0.0** | 0.0 | **0.0** | 0.0 | **0.0** | 0.0 | **0.0** | 0.013 | **0.013** |
| | $EB_R$ | -0.24 | -0.25 | 0.24 | 0.25 | -0.01 | **-0.01** | 0.0 | **0.0** | 0.49 | 0.5 | | |
| ItemKNN | $VB_I$ | 0.0 | **0.0** | 0.07 | **0.07** | -0.07 | **-0.07** | 0.0 | **0.0** | 0.14 | **0.14** | 0.052 | **0.052** |
| | $EB_I$ | -2.19 | **-2.02** | 2.27 | **2.1** | -0.07 | **-0.07** | -0.01 | **-0.01** | 4.55 | **4.21** | | |
| | $VB_R$ | 0.0 | **0.0** | 0.01 | **0.01** | -0.02 | **-0.02** | 0.0 | **0.0** | 0.03 | **0.03** | 0.052 | **0.052** |
| | $EB_R$ | -0.62 | **-0.61** | 0.64 | **0.64** | -0.02 | **-0.02** | -0.0 | **-0.0** | 1.29 | **1.27** | | |
| BPR | $VB_I$ | 0.14 | 0.25 | -0.11 | -0.22 | -0.03 | **-0.03** | 0.0 | **0.0** | 0.28 | 0.5 | 0.016 | **0.016** |
| | $EB_I$ | -1.5 | **-1.31** | 1.55 | **1.36** | -0.05 | **-0.05** | -0.0 | **-0.0** | 3.11 | **2.72** | | |
| | $VB_R$ | 0.0 | 0.01 | -0.01 | **-0.01** | 0.0 | **0.0** | 0.0 | **0.0** | 0.01 | 0.02 | 0.016 | **0.016** |
| | $EB_R$ | -0.05 | **-0.05** | 0.05 | **0.05** | -0.01 | **-0.01** | 0.0 | **0.0** | 0.11 | **0.11** | | |
| BiasedMF | $VB_I$ | 0.0 | **0.0** | 0.09 | **0.09** | -0.06 | **-0.06** | -0.04 | **-0.04** | 0.19 | **0.19** | 0.009 | **0.009** |
| | $EB_I$ | -3.41 | **-3.3** | 3.51 | **3.4** | -0.06 | **-0.06** | -0.04 | **-0.04** | 7.01 | **6.79** | | |
| | $VB_R$ | 0.0 | **0.0** | 0.02 | **0.02** | -0.0 | **-0.0** | -0.02 | **-0.02** | 0.04 | **0.04** | 0.009 | **0.009** |
| | $EB_R$ | -1.75 | **-1.73** | 1.78 | **1.75** | -0.01 | **-0.01** | -0.02 | **-0.02** | 3.55 | **3.51** | | |
| SVD++ | $VB_I$ | 0.0 | **0.0** | 0.02 | **0.02** | 0.0 | **0.0** | -0.02 | **-0.02** | 0.04 | **0.04** | 0.009 | **0.009** |
| | $EB_I$ | -3.4 | **-3.3** | 3.44 | **3.35** | -0.02 | **-0.02** | -0.02 | **-0.02** | 6.88 | **6.7** | | |
| | $VB_R$ | 0.0 | **0.0** | 0.0 | **0.0** | 0.0 | **0.0** | -0.0 | **-0.0** | 0.01 | **0.01** | 0.009 | **0.009** |
| | $EB_R$ | -1.74 | **-1.73** | 1.75 | **1.74** | -0.01 | **-0.01** | -0.01 | **-0.01** | 3.51 | **3.47** | | |

Table 5. The resulting popularity bias and accuracy of different mitigation approaches on MoviesLens-1M dataset.

| Algorithm | | g1 | | g2 | | g3 | | Total BS | | NDCG | |
|---|---|---|---|---|---|---|---|---|---|---|---|
| | | baseL. | MFAIR | baseL. | MFAIR | baseL. | MFAIR | baseL. | MFAIR | baseL. | MFAIR |
| MostPop | $VB_I$ | 34.43 | **34.4** | -28.6 | **-28.57** | -5.83 | **-5.83** | 68.85 | **68.8** | 0.104 | **0.104** |
| | $EB_I$ | 34.48 | **34.44** | -28.65 | **-28.61** | -5.83 | **-5.83** | 68.95 | **68.89** | | |
| | $VB_R$ | 34.48 | **34.4** | -28.65 | **-28.57** | -5.83 | **-5.83** | 68.96 | **68.8** | 0.106 | **0.106** |
| | $EB_R$ | 34.53 | **34.45** | -28.7 | **-28.62** | -5.83 | **-5.83** | 69.06 | **68.89** | | |
| RandomG | $VB_I$ | -35.05 | **-33.21** | 4.89 | **4.04** | 30.16 | **29.17** | 70.1 | **66.42** | 0.011 | **0.011** |
| | $EB_I$ | -35.2 | **-33.81** | 5.08 | **4.46** | 30.12 | **29.35** | 70.39 | **67.62** | | |
| | $VB_R$ | -33.78 | **-31.55** | 5.71 | **4.6** | 28.07 | **26.95** | 67.57 | **63.1** | 0.011 | **0.12** |
| | $EB_R$ | -33.91 | **-31.29** | 5.75 | **4.46** | 28.16 | **26.83** | 67.83 | **62.58** | | |
| UserKNN | $VB_I$ | 32.91 | **32.17** | -27.09 | **-26.35** | -5.82 | **-5.81** | 65.82 | **64.33** | 0.123 | **0.123** |
| | $EB_I$ | 33.46 | **32.9** | -27.64 | **-27.09** | -5.82 | **-5.81** | 66.91 | **65.8** | | |
| | $VB_R$ | 33.42 | **32.34** | -27.6 | **-26.52** | -5.82 | **-5.81** | 66.85 | **64.67** | 0.125 | **0.125** |
| | $EB_R$ | 33.82 | **33.03** | -28.0 | **-27.22** | -5.82 | **-5.82** | 67.64 | **66.07** | | |
| ItemKNN | $VB_I$ | 34.17 | **33.65** | -28.35 | **-27.82** | -5.83 | **-5.83** | 68.35 | **67.29** | 0.121 | **0.121** |
| | $EB_I$ | 34.28 | **33.89** | -28.45 | **-28.06** | -5.83 | **-5.83** | 68.56 | **67.77** | | |
| | $VB_R$ | 33.77 | **33.42** | -27.94 | **-27.6** | -5.83 | **-5.83** | 67.53 | **66.85** | 0.122 | **0.122** |
| | $EB_R$ | 33.96 | **33.7** | -28.13 | **-27.87** | -5.83 | **-5.83** | 67.91 | **67.4** | | |
| BPR | $VB_I$ | 18.12 | **12.23** | -13.52 | **-8.02** | -4.6 | **-4.21** | 36.25 | **24.46** | 0.137 | 0.135 |
| | $EB_I$ | 20.14 | **15.98** | -15.31 | **-11.44** | -4.83 | **-4.53** | 40.28 | **31.95** | | |
| | $VB_R$ | 19.65 | **18.83** | -14.69 | **-13.94** | -4.96 | **-4.88** | 39.29 | **37.65** | 0.141 | 0.14 |
| | $EB_R$ | 21.16 | **20.6** | -16.08 | **-15.57** | -5.08 | **-5.02** | 42.33 | **41.2** | | |
| BiasedMF | $VB_I$ | 6.28 | **3.07** | -5.55 | **-2.0** | -0.72 | -1.07 | 12.56 | **6.15** | 0.065 | 0.064 |
| | $EB_I$ | 8.62 | **5.63** | -7.55 | **-4.37** | -1.07 | -1.25 | 17.25 | **11.26** | | |
| | $VB_R$ | 6.98 | **4.32** | -6.05 | **-3.04** | -0.94 | -1.28 | 13.96 | **8.65** | 0.065 | **0.066** |
| | $EB_R$ | 8.9 | **6.37** | -8.07 | **-4.85** | -0.84 | -1.52 | 17.81 | **12.74** | | |
| SVD++ | $VB_I$ | 8.13 | **6.9** | -5.29 | **-4.22** | -2.83 | **-2.68** | 16.25 | **13.81** | 0.082 | 0.081 |
| | $EB_I$ | 11.4 | **10.16** | -8.25 | **-7.15** | -3.15 | **-3.01** | 22.8 | **20.32** | | |
| | $VB_R$ | 9.85 | **7.17** | -6.99 | **-4.43** | -2.86 | **-2.74** | 19.7 | **14.34** | 0.083 | **0.083** |
| | $EB_R$ | 12.67 | **10.65** | -9.57 | **-7.49** | -3.1 | -3.16 | 25.34 | **21.29** | | |

Table 6. The resulting popularity bias and accuracy of different mitigation approaches on Book-Crossing dataset.

| Algorithm | | g1 | | g2 | | g3 | | Total BS | | NDCG | |
|---|---|---|---|---|---|---|---|---|---|---|---|
| | | baseL. | MFAIR | baseL. | MFAIR | baseL. | MFAIR | baseL. | MFAIR | baseL. | MFAIR |
| MostPop | $VB_I$ | 34.1 | **34.1** | -25.01 | **-25.01** | -9.09 | **-9.09** | 68.2 | **68.2** | 0.019 | **0.019** |
| | $EB_I$ | 34.1 | **34.1** | -25.01 | **-25.01** | -9.09 | **-9.09** | 68.2 | **68.2** | | |
| | $VB_R$ | 34.1 | **34.1** | -25.01 | **-25.01** | -9.09 | **-9.09** | 68.2 | **68.2** | 0.019 | **0.019** |
| | $EB_R$ | 34.1 | **34.1** | -25.01 | **-25.01** | -9.09 | **-9.09** | 68.2 | **68.2** | | |
| RandomG | $VB_I$ | -32.26 | **-31.64** | 8.05 | **7.75** | 24.2 | **23.89** | 64.52 | **63.27** | 0.001 | **0.001** |
| | $EB_I$ | -32.21 | **-31.79** | 7.9 | **7.69** | 24.31 | **24.1** | 64.52 | **63.58** | | |
| | $VB_R$ | -31.6 | **-28.87** | 8.49 | **7.12** | 23.11 | **21.75** | 63.19 | **57.74** | 0.001 | **0.001** |
| | $EB_R$ | -31.62 | **-28.67** | 8.26 | **6.8** | 23.36 | **21.88** | 63.24 | **57.35** | | |
| UserKNN | $VB_I$ | 11.09 | **6.69** | -8.08 | **-4.73** | -3.01 | **-1.96** | 22.18 | **13.38** | 0.013 | **0.013** |
| | $EB_I$ | 12.51 | **9.54** | -9.6 | **-7.34** | -2.92 | **-2.2** | 25.02 | **19.08** | | |
| | $VB_R$ | 10.94 | **10.13** | -7.99 | **-7.37** | -2.95 | **-2.76** | 21.87 | **20.26** | 0.013 | **0.013** |
| | $EB_R$ | 12.43 | **11.88** | -9.56 | **-9.14** | -2.87 | **-2.75** | 24.86 | **23.76** | | |
| ItemKNN | $VB_I$ | 24.06 | **20.99** | -16.12 | **-13.51** | -7.94 | **-7.47** | 48.13 | **41.98** | 0.052 | **0.052** |
| | $EB_I$ | 25.07 | **22.97** | -16.99 | **-15.21** | -8.07 | **-7.75** | 50.13 | **45.93** | | |
| | $VB_R$ | 24.36 | **23.18** | -16.32 | **-15.31** | -8.04 | **-7.87** | 48.72 | **46.36** | 0.052 | **0.052** |
| | $EB_R$ | 25.26 | **24.48** | -17.12 | **-16.45** | -8.14 | **-8.03** | 50.53 | **48.96** | | |
| BPR | $VB_I$ | 16.62 | **13.67** | -10.09 | **-7.53** | -6.53 | **-6.13** | 33.24 | **27.33** | 0.016 | **0.016** |
| | $EB_I$ | 17.56 | **15.61** | -10.78 | **-9.09** | -6.78 | **-6.52** | 35.13 | **31.22** | | |
| | $VB_R$ | 16.68 | **16.6** | -10.07 | **-10.0** | -6.61 | **-6.6** | 33.36 | **33.19** | 0.016 | **0.016** |
| | $EB_R$ | 17.59 | **17.54** | -10.76 | **-10.71** | -6.83 | **-6.83** | 35.19 | **35.08** | | |
| BiasedMF | $VB_I$ | 33.94 | **33.06** | -24.99 | **-24.82** | -8.95 | **-8.24** | 67.89 | **66.12** | 0.009 | **0.009** |
| | $EB_I$ | 34.0 | **33.42** | -25.0 | **-24.89** | -9.0 | **-8.53** | 67.99 | **66.84** | | |
| | $VB_R$ | 34.05 | **33.8** | -24.97 | **-24.94** | -9.08 | **-8.86** | 68.1 | **67.61** | 0.009 | **0.009** |
| | $EB_R$ | 34.07 | **33.91** | -24.98 | **-24.97** | -9.09 | **-8.94** | 68.13 | **67.81** | | |
| SVD++ | $VB_I$ | 33.76 | **32.83** | -24.88 | **-24.58** | -8.88 | **-8.25** | 67.52 | **65.66** | 0.009 | **0.009** |
| | $EB_I$ | 33.78 | **33.27** | -24.93 | **-24.73** | -8.95 | **-8.54** | 67.75 | **66.54** | | |
| | $VB_R$ | 33.89 | **33.62** | -24.82 | **-24.75** | -9.07 | **-8.88** | 67.78 | **67.24** | 0.009 | **0.009** |
| | $EB_R$ | 33.96 | **33.79** | -24.89 | **-24.84** | -9.07 | **-8.95** | 67.92 | **67.57** | | |

## 5. Conclusions and future work

In this study, we addressed two types of bias in CF recommender systems at once: geographic bias and popularity bias. We presented MFAIR, a **m**ulti-**fa**cet post-processing b**i**as mitigation algo**r**ithm that simultaneously manages both biases in recommendations while maintaining accuracy. It considers the distribution of continents and popularity groups in the input data as the target distribution to achieve in the top-*k* recommendations. We applied MFAIR to the output of state-of-the-art CF recommender algorithms and compared our approach to a well-known post-processing method. The results showed that our approach better managed the trade-off between both types of bias and accuracy. It also performed better in terms of popularity bias in all experiments.

Recommender systems are prone to various types of bias depending on the industry in which they are used. When these biases emerge on an intersectional level, they can further harm sensitive groups. To address this problem, other aspects of fairness, such as user fairness based on gender, age, etc., or provider fairness based on providers' demographics, can be included in this study. Future work might involve examining how these different biases can be managed to provide optimal results for more of the involved stakeholders in the system.

## References

[1] F. Ricci, L. Rokach, and B. Shapira, "Recommender systems: introduction and challenges," in *Recommender systems handbook*: Springer, 2015, pp. 1-34.

[2] E. Gómez, C. Shui Zhang, L. Boratto, M. Salamó, and M. Marras, "The winner takes it all: geographic imbalance and provider (un) fairness in educational recommender systems," in *Proceedings of the 44th International ACM SIGIR Conference on Research and Development in Information Retrieval*, 2021, pp. 1808-1812.

[3] E. Gómez, L. Boratto, and M. Salamó, "Provider fairness across continents in collaborative recommender systems," *Information Processing & Management,* vol. 59, no. 1, p. 102719, 2022.

[4] B. Friedman and H. Nissenbaum, "Bias in computer systems," *ACM Transactions on information systems (TOIS),* vol. 14, no. 3, pp. 330-347, 1996.

[5] H. Abdollahpouri *et al.*, "Multistakeholder recommendation: Survey and research directions," *User Modeling and User-Adapted Interaction,* vol. 30, no. 1, pp. 127-158, 2020.

[6] D. Shakespeare, L. Porcaro, E. Gómez, and C. Castillo, "Exploring artist gender bias in music recommendation," *arXiv preprint arXiv:2009.01715,* 2020.

[7] A. Ferraro, X. Serra, and C. Bauer, "Break the loop: Gender imbalance in music recommenders," in *Proceedings of the 2021 Conference on Human Information Interaction and Retrieval*, 2021, pp. 249-254.

[8] M. Malek, "Criminal courts’ artificial intelligence: the way it reinforces bias and discrimination," *AI and Ethics,* vol. 2, no. 1, pp. 233-245, 2022.

[9] G. Farnadi, P. Kouki, S. K. Thompson, S. Srinivasan, and L. Getoor, "A fairness-aware hybrid recommender system," *arXiv preprint arXiv:1809.09030,* 2018.

[10] E. Gómez, L. Boratto, and M. Salamó, "Disparate impact in item recommendation: A case of geographic imbalance," in *European Conference on Information Retrieval*, 2021: Springer, pp. 190-206.

[11] M. Zehlike, K. Yang, and J. Stoyanovich, "Fairness in ranking: A survey," *arXiv preprint arXiv:2103.14000,* 2021.

[12] E. Gómez, C. S. Zhang, L. Boratto, M. Salamó, and G. Ramos, "Enabling cross-continent provider fairness in educational recommender systems," *Future Generation Computer Systems,* vol. 127, pp. 435-447, 2022.

[13] Y. Zhang *et al.*, "Causal intervention for leveraging popularity bias in recommendation," in *Proceedings of the 44th International ACM SIGIR Conference on Research and Development in Information Retrieval*, 2021, pp. 11-20.

[14] M. Zehlike, T. Sühr, R. Baeza-Yates, F. Bonchi, C. Castillo, and S. Hajian, "Fair Top-k Ranking with multiple protected groups," *Information Processing & Management,* vol. 59, no. 1, p. 102707, 2022.

[15] A. B. Ahanger, S. W. Aalam, M. R. Bhat, and A. Assad, "Popularity Bias in Recommender Systems-A Review," in *International Conference on Emerging Technologies in Computer Engineering*, 2022: Springer, pp. 431-444.

[16] N. Ranjbar Kermany, W. Zhao, J. Yang, J. Wu, and L. Pizzato, "A fairness-aware multi-stakeholder recommender system," *World Wide Web,* vol. 24, no. 6, pp. 1995-2018, 2021.

[17] E. Walster, E. Berscheid, and G. W. Walster, "New directions in equity research," *Journal of personality and social psychology,* vol. 25, no. 2, p. 151, 1973.

[18] N. R. Kermany, W. Zhao, J. Yang, and J. Wu, "Reincre: Enhancing collaborative filtering recommendations by incorporating user rating credibility," in *International Conference on Web Information Systems Engineering*, 2020: Springer, pp. 64-72.

[19] J. Lu, D. Wu, M. Mao, W. Wang, and G. Zhang, "Recommender system application developments: a survey," *Decision Support Systems,* vol. 74, pp. 12-32, 2015.

[20] S. Reddy, S. Nalluri, S. Kunisetti, S. Ashok, and B. Venkatesh, "Content-based movie recommendation system using genre correlation," in *Smart Intelligent Computing and Applications*: Springer, 2019, pp. 391-397.

[21] L. Jiang, Y. Cheng, L. Yang, J. Li, H. Yan, and X. Wang, "A trust-based collaborative filtering algorithm for E-commerce recommendation system," *Journal of ambient intelligence and humanized computing,* vol. 10, no. 8, pp. 3023-3034, 2019.

[22] X. Yu, F. Jiang, J. Du, and D. Gong, "A cross-domain collaborative filtering algorithm with expanding user and item features via the latent factor space of auxiliary domains," *Pattern Recognition,* vol. 94, pp. 96-109, 2019.

[23] R. Sujithra Alias Kanmani, B. Surendiran, and S. Ibrahim, "Recency augmented hybrid collaborative movie recommendation system," *International Journal of Information Technology,* vol. 13, no. 5, pp. 1829-1836, 2021.

[24] R. Burke, N. Sonboli, M. Mansoury, and A. Ordoñez-Gauger, "Balanced neighborhoods for fairness-aware collaborative recommendation," 2017.

[25] Y. Li, H. Chen, Z. Fu, Y. Ge, and Y. Zhang, "User-oriented fairness in recommendation," in *Proceedings of the Web Conference 2021*, 2021, pp. 624-632.

[26] H. Abdollahpouri, M. Mansoury, R. Burke, and B. Mobasher, "The unfairness of popularity bias in recommendation," *arXiv preprint arXiv:1907.13286,* 2019.

[27] T. Qi *et al.*, "ProFairRec: Provider Fairness-aware News Recommendation," *arXiv preprint arXiv:2204.04724,* 2022.

[28] H. Abdollahpouri, R. Burke, and B. Mobasher, "Controlling popularity bias in learning-to-rank recommendation," in *Proceedings of the eleventh ACM conference on recommender systems*, 2017, pp. 42-46.

[29] S. Lin, J. Wang, Z. Zhu, and J. Caverlee, "Quantifying and Mitigating Popularity Bias in Conversational Recommender Systems," *arXiv preprint arXiv:2208.03298,* 2022.

[30] H. Wu, C. Ma, B. Mitra, F. Diaz, and X. Liu, "Multi-FR: A Multi-Objective Optimization Method for Achieving Two-sided Fairness in E-commerce Recommendation," *arXiv preprint arXiv:2105.02951,* 2021.

[31] G. K. Patro, A. Biswas, N. Ganguly, K. P. Gummadi, and A. Chakraborty, "Fairrec: Two-sided fairness for personalized recommendations in two-sided platforms," in *Proceedings of The Web Conference 2020*, 2020, pp. 1194-1204.

[32] H. Abdollahpouri, M. Mansoury, R. Burke, B. Mobasher, and E. Malthouse, "User-centered evaluation of popularity bias in recommender systems," in *Proceedings of the 29th ACM Conference on User Modeling, Adaptation and Personalization*, 2021, pp. 119-129.

[33] M. Zehlike and C. Castillo, "Reducing disparate exposure in ranking: A learning to rank approach," in *Proceedings of The Web Conference 2020*, 2020, pp. 2849-2855.

[34] C.-N. Ziegler, S. M. McNee, J. A. Konstan, and G. Lausen, "Improving recommendation lists through topic diversification," in *Proceedings of the 14th international conference on World Wide Web*, 2005, pp. 22-32.

[35] K. Järvelin and J. Kekäläinen, "Cumulated gain-based evaluation of IR techniques," *ACM Transactions on Information Systems (TOIS),* vol. 20, no. 4, pp. 422-446, 2002.